\documentclass[traditabstract]{aa} % for the abstract without structuration 
\usepackage{natbib}
\usepackage{graphicx}

\newcommand{\new}[1]{{#1}}

\usepackage{txfonts}
\begin{document}
   \title{AAOmega spectroscopy of 29\,351 stars in fields centered on ten
  Galactic globular clusters\thanks{The data described in Tables 1, 2 \& 3 are
    only available in electronic form at the CDS via anonymous ftp to {\tt
      cdsarc.u-strasbg.fr (130.79.128.5)} or via {\tt
      http://cdsweb.u-strasbg.fr/cgi-bin/qcat?J/A+A/530/A31}}}

   \subtitle{}

   \author{Richard R. Lane\inst{1,2}\fnmsep\thanks{rlane@astro-udec.cl},
     L\'aszl\'o L. Kiss\inst{2,3}, Geraint F. Lewis\inst{2}, Rodrigo
     A. Ibata\inst{4}, Arnaud Siebert\inst{4}, Timothy R. Bedding\inst{2},
     P\'eter Sz\'ekely\inst{5} \and Gyula M. Szab\'o\inst{3} }

   \institute{Departamento de Astronom\'ia, Universidad de Concepci\'on, Casilla 160 C, Concepci\'on, Chile
\and
Sydney Institute for Astronomy, School of Physics,
A28, The University of Sydney, NSW, Australia 2006
\and
Konkoly Observatory of the Hungarian Academy of Sciences, PO Box 67, H-1525, Budapest, Hungary
\and
Observatoire Astronomique, Universite de Strasbourg, CNRS, 67000 Strasbourg, France
\and
Department of Experimental Physics, University of Szeged, Szeged 6720, Hungary
}

   \date{Accepted April 4, 2011}

 \abstract{Galactic globular clusters have been pivotal in our understanding
   of many astrophysical phenomena. Here we publish the extracted stellar
   parameters from a recent large spectroscopic survey of ten globular
   clusters. A brief review of the project is also presented. Stellar
   parameters have been extracted from individual stellar spectra using both a
   modified version of the Radial Velocity Experiment (RAVE) pipeline and a
   pipeline based on the parameter estimation method of RAVE. We publish here
   all parameters extracted from both pipelines. We calibrate the metallicity
   and convert this to [Fe/H] for each star and, furthermore, we compare the
   velocities and velocity dispersions of the Galactic stars in each field to
   the Besan\c{c}on Galaxy model. We find that the model does not correspond
   well with the data, indicating that the model is probably of little use for
   comparisons with pencil beam survey data such as this.}{}{}{}{}

% \abstract{}{Galactic globular clusters have been pivotal in our understanding
%   of many astrophysical phenomena. Here we publish the extracted stellar
%   parameters from a recent large spectroscopic survey of ten globular
%   clusters. A brief review of the project is also presented.}{Stellar
%   parameters have been extracted from individual stellar spectra using both a
%   modified version of the Radial Velocity Experiment (RAVE) pipeline and a
%   pipeline based on the parameter estimation method of RAVE. We publish here
%   all parameters extracted from both pipelines.}{We calibrate the metallicity
%   and convert this to [Fe/H] for each star and, furthermore, we compare the
%   velocities and velocity dispersions of the Galactic stars in each field to
%   the Besan\c{c}on Galaxy model. We find that the model does not correspond
%   well with the data, indicating that the model is probably of little use for
%   comparisons with pencil beam survey data such as this.}{}

   
   \authorrunning{Lane, et al.}
   \titlerunning{}

   \maketitle
%
%________________________________________________________________

\section{Introduction}

Globular clusters (GCs) are intriguing astronomical objects for many
reasons. They are invariably found surrounding spiral and elliptical
galaxies, have been used as tracers of galactic potentials
\cite[e.g.][]{Kissler-Patig99,Gebhardt09} and can be employed to test
gravitational theories \cite[e.g.][]{Lane09,Sollima10}. Despite
decades of detailed examination, it is still uncertain how GCs formed
\cite[][]{Bekki02,Lipscy04,Mashchenko05,Griffen10,Lee10} and their
dark matter (DM) content is still debated, although they are usually
considered to be DM-poor \cite[][and references
therein]{Lane09}. Recently, several studies revealed an interesting
possible flattening in the velocity dispersions of several GCs at
large radii, reminiscent of large elliptical galaxies, a signature of
the kinematics of DM-dominated objects \cite[][]{Scarpa07}. These
results called into question either the paucity of DM in GCs or our
understanding of the nature of the gravitational interaction at low
accelerations (below $a_0\sim10^{-10}$\,ms$^{-2}$).

We performed detailed kinematic studies on ten Galactic GCs, with the goal of
calculating independent velocity dispersion profiles to determine whether
these results could be replicated \cite[][hereafter Papers I, II and III
  respectively]{Lane09,Lane10a,Lane10b}. In short, we found that the
flattening of the velocity dispersion profiles shown in previous studies were
not reproducible, despite two of our clusters being chosen to overlap with
earlier investigations, namely M30 (Paper I) and NGC 288 (Paper III). Our
results indicated that neither DM, nor a modification of gravitational theory,
were required to reconcile the observed kinematics of our target GCs with
current theory. In addition, we extended a recent metallicity calibration
technique for open and globular clusters using the equivalent widths of the
calcium triplet lines and the horizontal branch magnitude of the cluster
\citep{Cole04,Warren09}, to the $K$ band magnitude of the tip of the Red Giant
Branch. This is several magnitudes brighter and can, therefore, be used for
more distant clusters, and, more importantly, for GCs with hot, blue,
horizontal branches whose stars do not exhibit strong calcium triplet lines
(Section \ref{params}). Furthermore, a broad measure of the strength of the
Galactic tidal field at various Galactocentric distances was made by comparing
the velocity dispersions of the external and internal parts of the GCs. On the
basis that the member stars at large radii are affected more by the external
gravitational field, we showed that any flattening of the velocity dispersion
profile could be attributed to this external field and did not require DM or
modifications to gravity. We also argued that the lack of tidal heating
signatures in our GCs at large Galactocentric radii is weakly suggestive of a
spherical dark Halo (Paper III). In addition, we discovered the exciting
possibility of a two-component kinematic population within 47 Tucanae, which
we interpreted as evidence for the cluster forming as two individual clumps in
the protocluster cloud which coalesced at a later date \cite[with an upper
  limit of $\sim7.3\pm1.5$ Gyr ago;][]{Lane10c}. We also found an as yet
unexplained anomalously rapid cooling of the outer regions of GCs following
tidal heating by the Galactic disc.

From these surveys we produced the largest sample to date of spectral data of
both \object{47 Tucanae} (\object{47 Tuc}) and \object{M55} members, as well
as large numbers of members of \object{M4}, \object{M12}, \object{M22},
\object{M30}, \object{M53}, \object{M68}, \object{NGC 288} and \object{NGC
  6752}. Although the main goal of this manuscript is to make these data
public for further research, from these surveys we also obtained spectra from
Galactic field stars in the foreground of the GCs, therefore, we also have a
large sample of Galactic spectra at various latitudes and longitudes. In the
current paper we compare the velocity dispersions of the Galactic stars at
each location to current dynamical models of the Milky Way. Projects such as
HERMES\footnote{{\tt http://www.aao.gov.au/AAO/HERMES/}} \cite[High Resolution
  Multi-object Echelle Spectrograph;][]{Barden08} on the Anglo-Australian
Telescope show the importance of radial velocity/metallicity surveys over a
large range of Galactic coordinates (particularly moving from the Disc to the
Halo) for understanding Galactic evolution. Here we present our pencil beam
survey as an analysis of the Galactic velocity dispersion as a precursor to
surveys such as HERMES. Furthermore, a comparison is made between
metallicities calculated using $\chi^2$ analysis, as used by the Radial
Velocity Experiment \cite[RAVE;][]{Zwitter08}, and the equivalent width method
used by \cite{Lane10a}. Because our sample of stars from the 47 Tuc field is
by far the largest, and also contains two GCs from the Small Magellanic Cloud
(NGC 121 and Kron 3), we have used this field for most discussions and figures
in the current paper.

%__________________________________________________________________

\section{Derived Parameters}\label{params}

All data was obtained on the AAOmega instrument on the Anglo-Australian
Telescope (AAT), with the same gratings.  We used the 1700D grating
($R=10\,000$) on the red arm to observe the calcium triplet region
($\sim8340$\AA$-8840$\AA) and the 1500V grating ($R=3700$) on the blue arm to
include the swathe of iron and magnesium lines around $\sim5200$\AA. A
detailed description of the observations, the membership selection process,
and other details of the programme were given in Papers I, II and III.

Our data pipeline \cite[see][]{Kiss07}, which calculates various stellar
parameters, is based on the RAVE pipeline. All data from this programme was
reduced using both the RAVE pipeline (which has been modified for use with
AAOmega data) and our own for comparison. These two pipelines work in slightly
different, albeit similar, manners. The \cite{Kiss07} pipeline uses an
iterative process to obtain best fits to synthetic spectra from the library by
\cite{Munari05}, which are degraded to the resolution of AAOmega, and this
model is cross-correlated with the observed spectra to calculate the stellar
parameters. The RAVE method obtains the best fit templates using penalised
$\chi^2$ rather than cross correlation \cite[a detailed description of the
  template matching, and subsequent parameter estimation processes for each
  are given by][]{Kiss07,Zwitter08}. The \cite{Kiss07} version of the pipeline
also differs slightly from the modified RAVE pipeline in that it trilinearly
interpolates the spectra in the synthetic library (refining the grid in
$T_{\rm eff}$, log\,$g$ and [m/H]). This leads to resolution of 50\,K in
$T_{\rm eff}$ , 0.1 in log\,$g$ and 0.1 in [m/H], whereas the RAVE version
reduces to a nonlinear interpolation on the six dimensions of the parameter
space. Furthermore, The \cite{Kiss07} pipeline enforces one more iteration
than the RAVE pipeline in the fitting process when calculating the radial
velocities, which may provide a small improvment in the quality of the radial
velocity estimate.

\begin{figure}[!ht]
  \begin{centering}
  \includegraphics[angle=-90,width=0.48\textwidth]{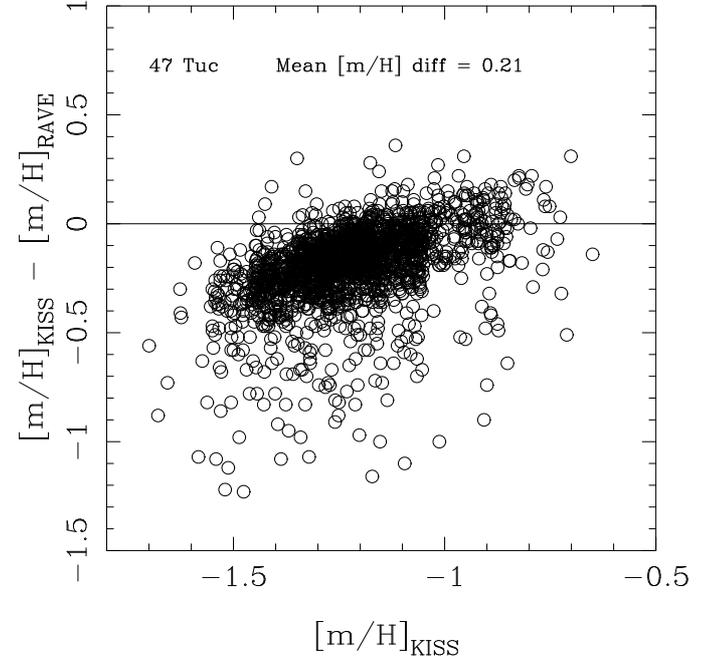}
  \caption{Comparison of [m/H] from the RAVE pipeline versus that from
  our own for all stars in the 47 Tuc field. Notice the small offset between
  the parameter values.}
  \label{mH}
  \end{centering}
\end{figure}

In most cases the RAVE extracted parameters are very similar to those from our
own pipeline, however, there is a small but significant offset between the
uncalibrated metallicity ([m/H]) extracted from the RAVE pipeline and those
from ours (Figure \ref{mH}). This is a known limitation of the RAVE pipeline
which overestimates the metallicity for low metallicities due to the noise
model assuming the same S/N for all pixels \citep{Zwitter08}. Due to these
subtle differences between the methods, and parameter estimates, we publish
here all parameters from both pipelines. Published parameters are shown in
Tables \ref{paramstbl1}, \ref{paramstbl2} and \ref{paramstbl3}; all data are
available via the CDS.

\new{\subsection{Parameter Uncertainties}

JHK magnitudes are taken directly from the 2 Micron All Sky Survey (2MASS),
and, therefore, have an uncertainty of $\sim0.03$ magnitudes
\cite[][]{Skrutskie06}. V and I magnitudes are estimated from 2MASS JHK using
unpublished transformations by G. Bakos (private communication). We have
verified that these transformations have an uncertainty of $\rm V\sim0.2$ and
$\rm I\sim0.1$ magnitudes by cross correlation with data by \cite{Weldrake04},
who provide VI photometry of 43\,067 stars within $\sim30'$ of 47 Tuc to
better than 0.03 magnitudes.

Rotational velocity ($v_{\rm rot}$) estimates are theoretical because they
come directly from the synthetic spectra by \cite{Munari05}, however, the lack
of any inclination information means $v_{\rm rot}$ can be considered
$v\cdot$sin($i$). We do not derive a formal uncertainy on the quoted values of
$v_{\rm rot}$, but because the two pipelines derive $v_{\rm rot}$ directly
from the synthetic spectra, we can estimate an uncertainty by simple
comparison between the two pipelines. The mean difference of $v_{\rm rot}$
between the two pipelines is $\sim12$\,km\,s$^{-1}$. Taken at face value this
can be considered the uncertainty on $v_{\rm rot}$, however, due to the
spectral resolution of the observations we advise that values of $v_{\rm
  rot}\lesssim30$\,km\,s$^{-1}$ are much less reliable.

Similarly, because $T_{\rm eff}$ and log\,$g$ are taken directly from the
synthetic spectra for both pipelines, and because we do not derive any formal
uncertainties for these parameters, the mean differences between the two
pipelines (276.5\,K and 0.6, respectively, see Figure \ref{tefflogg}) can
be taken as the face value uncertainties.

At the resolution of our observations, and due to the density of the template
spectra, microturbulence values of $\lesssim2$\,km\,s$^{-1}$ are not
resolved. Therefore we recommend that this be taken as the minimum uncertainty
on the microturbulence values, although the true uncertainties are likely to
be larger.

The uncertainties in all metallicities ([m/H], [M/H] and [Fe/H]) are
$\pm0.1$ dex (see Section \ref{metalcalib} and Figure \ref{metals}).

}

\begin{table}[!h]
\caption{Param\new{e}ters published in the current paper from the RAVE
  pipeline for all fields. The final column designates whether the star was
  classified as a member in Papers I, II and III. \new{See text for
    explanations of the parameters and associated
    uncertainties.}}\label{paramstbl1}
\begin{tabular}{@{}l@{}}
\hline
\hline
Cluster/Field\\
RA (radians)\\
dec (radians)\\
estimated I magnitude\\
field name\\
fibre number\\
$V_{\rm r}$ (km\,s$^{-1}$)\\
$V_{\rm r}$ uncertainty (km\,s$^{-1}$)\\
$T_{\rm eff}$ (K)\\
log\,$g$\\
$\rm{[m/H]}$\\
$\rm{[\alpha/Fe]}$\\
microturbulence \new{(km\,s$^{-1}$)}\\
rotational velocity \new{($v\cdot$sin[$i$])}\\
cluster member (yes/no)\\
\hline
\hline
\end{tabular}
\end{table}

\begin{table}[!h]
\caption{Param\new{e}ters published in the current paper from the Kiss
  pipeline (with fields centered on M22, M30, \new{M53 and M68}). The final
  column designates whether the star was classified as a member in Papers I,
  II and III. \new{See text for
    explanations of the parameters and associated
    uncertainties.}}\label{paramstbl2}
\begin{tabular}{@{}l@{}}
\hline
\hline
Cluster/Field\\
Star ID\\
$V_{\rm r}$ (km\,s$^{-1}$)\\
$V_{\rm r}$ uncertainty (km\,s$^{-1}$)\\
sum of the equivalent widths of the CaT lines (\AA)\\
RA (degrees)\\
dec (degrees)\\
estimated V magnitude\\
$T_{\rm eff}$ (K)\\
log\,$g$\\
$\rm{[m/H]}$\\
rotational velocity \new{($v\cdot$sin[$i$])}\\
distance from cluster centre ($''$)\\
position angle\\
cluster member (yes/no)\\
\hline
\hline
\end{tabular}
\end{table}

\begin{table}[!h]
\caption{Param\new{e}ters published in the current paper from the Kiss
  pipeline (with fields centered on 47 Tuc, M12, M4, M55, NGC 288, NGC
  6752). The final column designates whether the star was classified as a
  member in Papers I, II and III. \new{See text for
    explanations of the parameters and associated
    uncertainties.}}\label{paramstbl3}
\begin{tabular}{@{}l@{}}
\hline
\hline
Cluster/Field\\
Star ID\\
$V_{\rm r}$ (km\,s$^{-1}$)\\
$V_{\rm r}$ uncertainty (km\,s$^{-1}$)\\
Sum of the equivalent widths of the CaT lines (\AA)\\
RA (degrees)\\
dec (degrees)\\
estimated I magnitude\\
estimated V magnitude\\
$T_{\rm eff}$ (K)\\
log\,$g$\\
$\rm{[m/H]}$\\
rotational velocity \new{($v\cdot$sin[$i$])}\\
J mag\\
H mag\\
K mag\\
distance from cluster centre ($''$)\\
position angle\\
cluster member (yes/no)\\
\hline
\hline
\end{tabular}
\end{table}

\begin{figure}[!ht]
  \begin{centering}
  \includegraphics[angle=-90,width=0.48\textwidth]{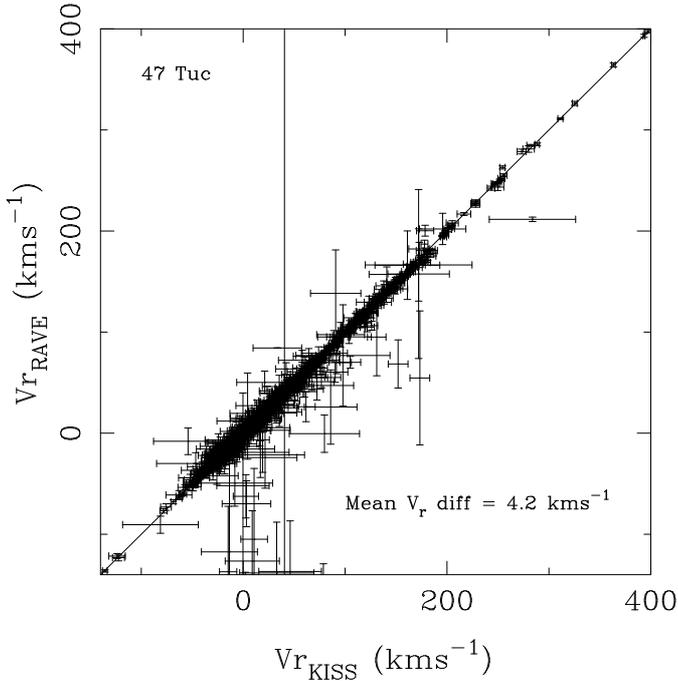}
  \caption{Comparison of radial velocity estimates from the RAVE pipeline
    versus that from our own for all the stars in the 47 Tuc field. Comparing
    only those stars determined to be members, the mean difference reduces to
    $\sim0.3$\,km\,s$^{-1}$ (Paper II). This is due to hot stars having strong
    Paschen lines (see text).}
  \label{vel}
  \end{centering}
\end{figure}

\begin{figure}[!ht]
  \begin{centering}
  \includegraphics[angle=-90,width=0.48\textwidth]{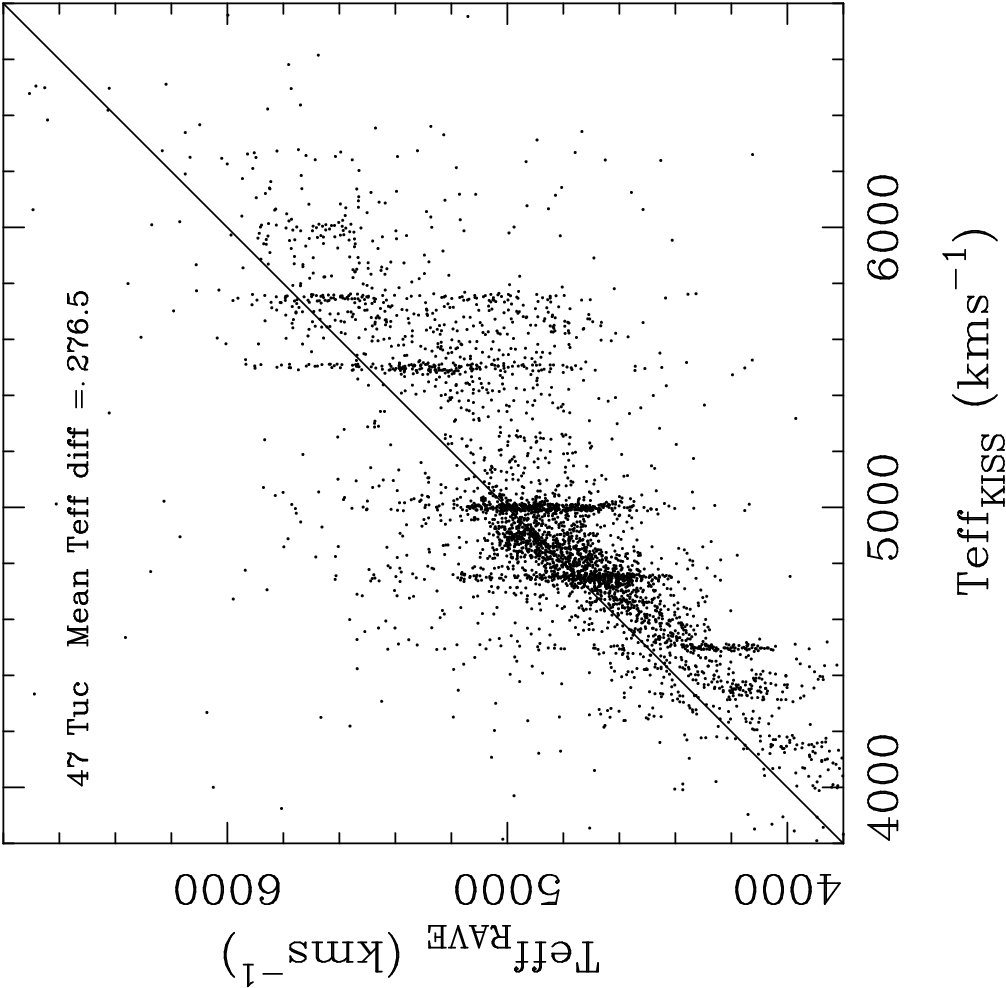}
  \includegraphics[angle=-90,width=0.48\textwidth]{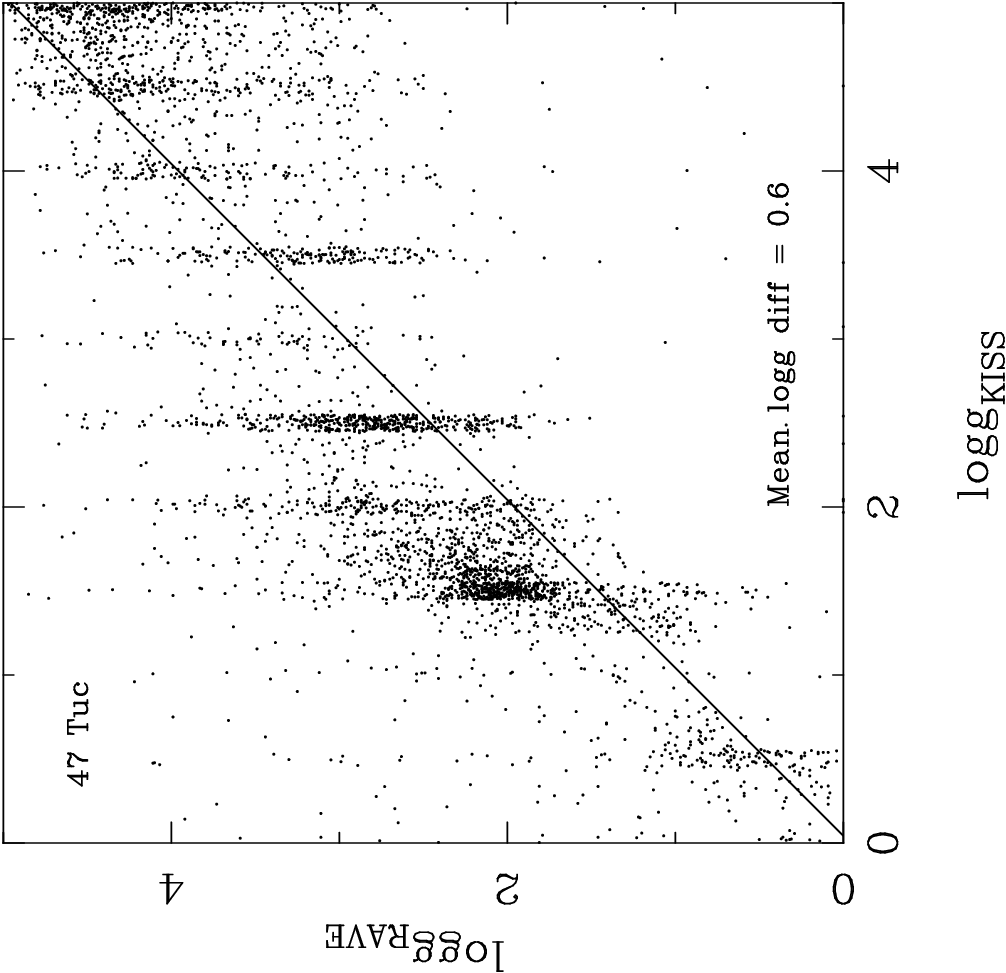}
  \caption{Comparisons between the estimated $T_{\rm eff}$ and
  log\,$g$ from each of the pipelines. Note that there is no obvious
  systematic offset between the two and that the average difference is
  small.}
  \label{tefflogg}
  \end{centering}
\end{figure}

Note that \new{both} [$\alpha$/Fe] and microturbulence estimates are
highly unreliable \new{and should be treated with caution} \cite[see][for
  \new{additional} details of parameter reliability and uncertainties for
  parameter estimates from the RAVE pipeline]{Zwitter08}. Furthermore, the
velocity uncertainties from the RAVE pipeline are the errors on the fits of
the maxima of the correlation functions using a quadratic function. They are,
therefore, not physical uncertainties on the radial velocities and tend to
overerestimate the true uncertainties by about 20\%. The radial velocity
uncertainties from the Kiss pipeline are the formal errors from gaussian fits
to the cross correlation function profiles. Again, these are not physical
uncertainties and may also slightly overestimate the true uncertainties,
however, the difference between the velocity estimates themselves from the
different pipelines is small. Although the overall mean difference between the
two pipelines for the 47 Tuc field is $\sim 4.2$\,km\,s$^{-1}$, this is mainly
due to the large uncertainties in radial velocity estimates for the hot stars
\new{and reduces to $\sim0.3$\,km\,s$^{-1}$ when comparing only those stars
  selected as members} (see Figure \ref{vel}); derived velocities for stars
with $T_{\rm eff}\gtrsim9000$\,K are much less reliable as these have the CaT
lines overtaken by strong P13, P15, and P16 hydrogen Paschen lines
\cite[e.g.][]{Fremat96}.

The RAVE collaboration represents the state of the art in the extraction of
many parameters from stellar spectra. The small deviations between the
parameters extracted with the modified RAVE pipeline and those extracted using
our own pipeline show that our software is also representative at this level
(Figures \ref{vel} and \ref{tefflogg}).

\section{Results}

\subsection{Kinematics of Galactic Field Stars}\label{kinematics}

\begin{figure}[!ht]
  \begin{centering}
  \includegraphics[angle=-90,width=0.48\textwidth]{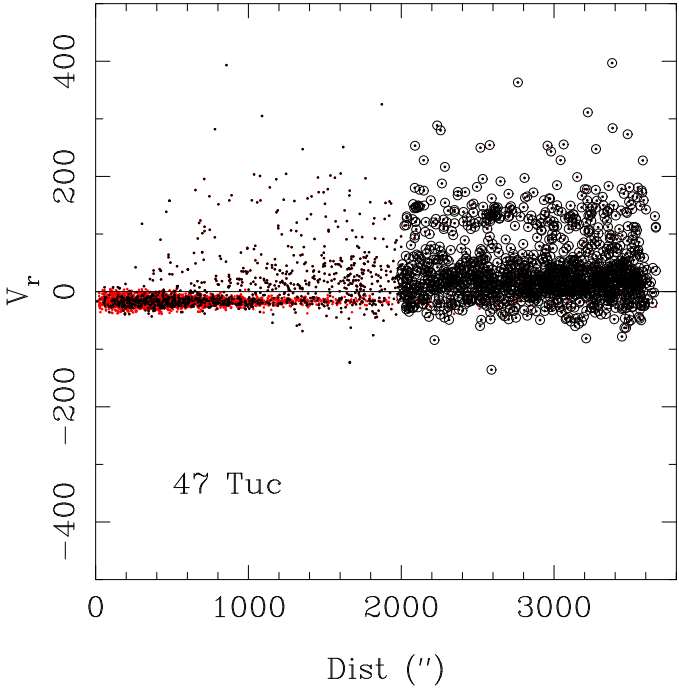}
  \caption{Stars observed within in the 2dF field centered on 47 Tuc. Red
    points indicate those stars selected as members in Paper II and black
    points indicate those selected as non-members. Note the overdensity of
    black points within the velocity range of the cluster which are likely to
    be members. \new{Circled black points indicate stars selected as
      non-members for the current paper.}}
  \label{members}
  \end{centering}
\end{figure}

Despite the main focus of this paper being the publication of the
data, we include some analysis of the field stars here. To do this, it
was first necessary to exclude all stars considered members of the GCs
(i.e. extract the field stars from the complete dataset). Our strict
membership selection method in Papers I, II and III evidently did not
extract all cluster members because it is obvious some have been left
behind (Figure \ref{members}). It was, therefore, necessary to impose
an additional cut based on the distance from the centre of the cluster
to ensure we removed all cluster members before the model comparisons.

Comparing our results with those of the best available dynamic Galaxy model
\cite[the Besan\c{c}on model;][see also {\tt
    http://model.obs-besancon.fr/}]{Robin03} gives a measure of the accuracy
of the model, which is known to have limitations \cite[e.g. the Disc component
  truncates at 10\,kpc, also see Figure 8 by][]{Conn08}. \new{Table
  \ref{modelvsdata}} shows the velocity dispersion and mean velocities of all
non-members from our ten fields and the Besan\c{c}on model. The model fields
have the same field centres as our data fields and were chosen to contain a
similar number of stars, with the same temperature ranges as our data. The
velocity dispersions for the data were calculated using the Markov Chain Monte
Carlo method described in Paper III. The lack of agreement between the model
and our data is apparent in \new{Table \ref{modelvsdata}}, however, the
Besan\c{c}on model is regularly employed for comparison with observational
surveys \cite[e.g.][]{Conn05,Conn07,Conn08}. The deviation of the model from
the data in the current paper highlights the model's limitations for
comparisons with {\it pencil beam} surveys. This should be taken into
consideration when using this model for analysis. The Besan\c{c}on model is,
however, very effective when used for comparisons with {\it large area}
surveys \cite[e.g.][]{Reyle10}.

A possible source of the model's discrepancies with small field surveys is
that, while the model does include the warp and flare of the Disc, it does not
include any known tidal streams, of which there are many
\cite[e.g.][]{Belokurov07}. Furthermore, note an apparent kinematically
distinct population of stars with $100\lesssim V_{\rm
  r}\lesssim200$\,km\,s$^{-1}$ in the 47 Tuc field (Figure \ref{members}). The
Small Magellanic Cloud is in this field (Paper II) which also has a
recessional velocity of $100\lesssim V_{\rm r}\lesssim200$\,km\,s$^{-1}$
\cite[e.g.][]{Storm04,Evans08,DePropis10}. The model cannot be expected to
replicate the mean velocity, or dispersion, of a field such as this.

\begin{table}[!h]
\caption{\new{Comaprison between the velocity dispersion ($\sigma$) and systemic
  (mean) velocity of non-cluster member stars in each field and the
  Besan\c{c}on model fields. All values are given in
  km\,s$^{-1}$.}}\label{modelvsdata}
\begin{tabular}{@{}|c|cc|cc|@{}}
\hline
Cluster/field & $\sigma_{\rm data}$ & $\sigma_{\rm model}$ & $V_{\rm sys(data)}$ &
$V_{\rm sys (model)}$ \\
\hline
M4 & 61.7 & 78.2 & -15.1 & -36.2 \\
M12 & 53.2 & 68.1 & -20.4 & 0.9 \\
M22 & 60.9 & 81.3 & -2.8 & 32.6 \\
M30 & 49.0 & 52.6 & -10.9 & -9.4 \\
M53 & 53.0 & 40.0 & 0.1 & -2.6 \\
M55 & 59.4 & 74.6 & -8.8 & 7.6 \\
M68 & 35.9 & 51.4 & 7.2 & 12.8 \\
NGC 288 & 40.8 & 38.1 & 7.8 & 6.1 \\
NGC 6752 & 50.4 & 62.6 & -5.5 & -15.5 \\
47 Tuc & 58.8 & 49.8 & 39.1 & 16.0 \\
\hline
\end{tabular}
\end{table}

%\begin{figure*}[!ht]
%  \begin{centering}
%  \includegraphics[angle=-90,width=0.67\textwidth]{figures/sysvel+veldisp_withB%esancon2.ps}
%  \caption{Velocity dispersion and mean (systemic) velocity for all
%  non-cluster member stars in each field and the Besan\c{c}on model of
%  each field for comparison. The points with error bars represent the
%  data and those without are the model. Straight lines connect data
%  and model points of the same field.}
%  \label{modelvsdata}
%  \end{centering}
%\end{figure*}

\subsection{Metallicity Calibration Based on [m/H], $T_{\rm eff}$ and
  ${\rm log}\,g$}\label{metalcalib}

\begin{figure}[!ht]
  \begin{centering}
  \includegraphics[angle=-90,width=0.48\textwidth]{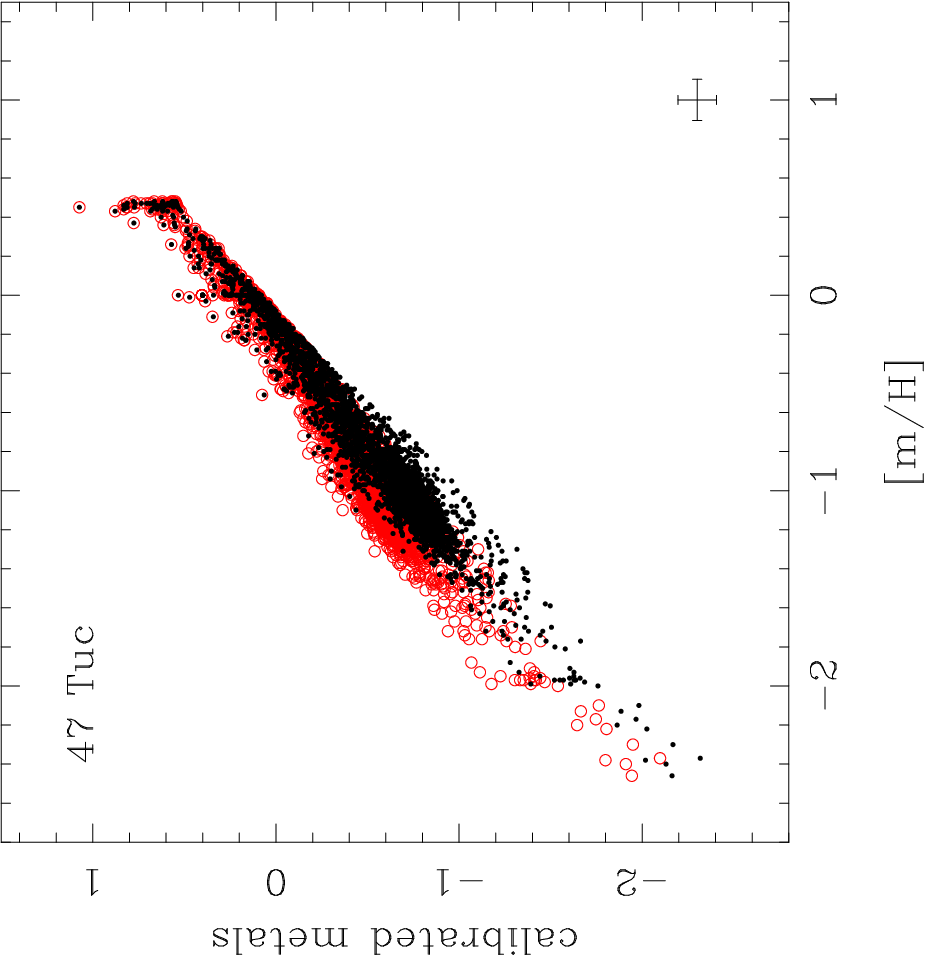}
  \includegraphics[angle=-90,width=0.48\textwidth]{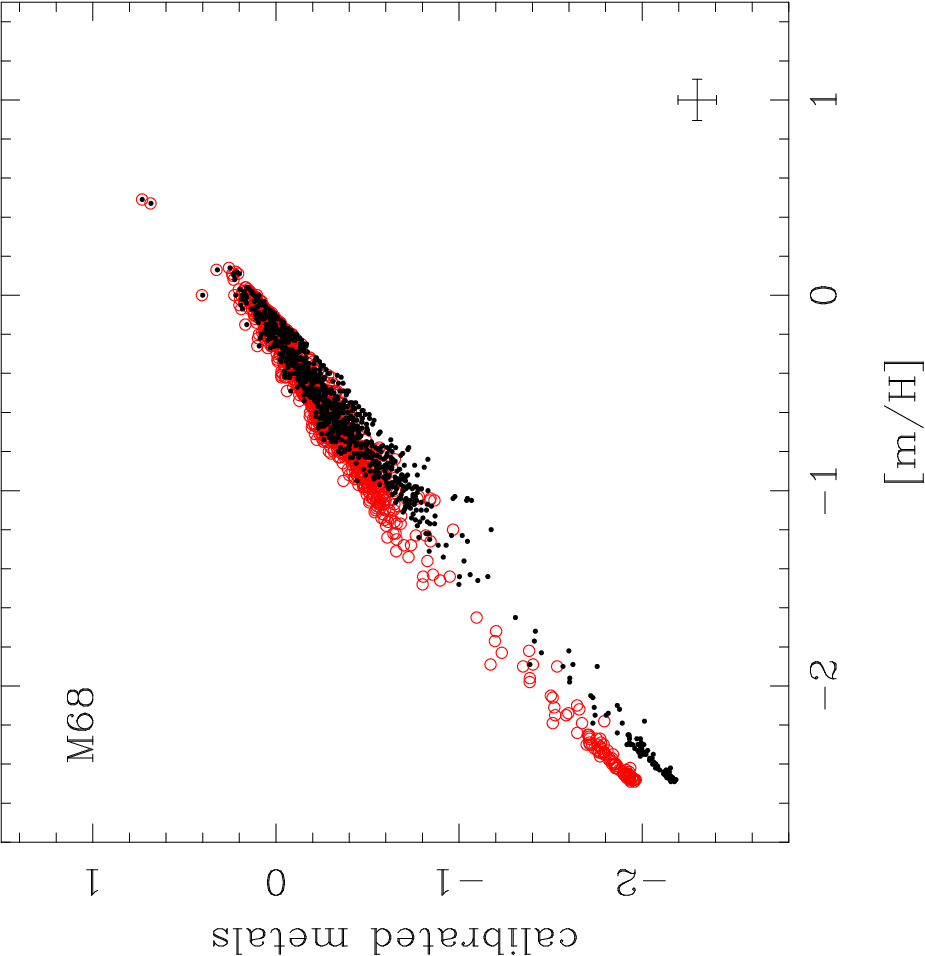}
  \caption{The [Fe/H] and [M/H] calibrations for all stars in the 47 Tuc and
    M68 fields. These match well with the values of [Fe/H] calculated in
    Papers II \& III via the equivalent widths of the CaII triplet lines and
    $K$ magnitude of the Tip of the Red Giant Branch. The red circles
    represent [M/H] as the ordinate, and the black \new{points} represent
    \new{[Fe/H]} as the ordinate. \new{A representative error bar is shown in
      the lower right of each panel.}}
  \label{metals}
  \end{centering}
\end{figure}

The RAVE project calibrated [M/H] and [Fe/H] based on [m/H], [$\alpha$/Fe] and
${\rm log}\,g$ \cite[see Equations 19, 20 and 21 by][who found that, using
  this method, the results for both {[}M/H{]} and {[}Fe/H{]} were within
  $\sim0.2$ dex of reference metallicities]{Zwitter08}. We have used this same
method to calculate [M/H] and [Fe/H] for all stars in the current
survey. Since our pipeline does not calculate [$\alpha$/Fe] we cannot directly
compare [Fe/H] calibrations based on our pipeline to those using the modified
RAVE pipeline. However, [Fe/H] was calculated for the ten members of NGC 121
found in the 47 Tuc field in Paper II. These ten stars can also be seen in
Figure \ref{metals} as the overdensity at [Fe/H]$\sim-1.6$. In Paper II the
metallicity for these stars were calculated using an equivalent width
calibration method which found [Fe/H]$=-1.50\pm0.10$ for NGC 121, completely
consistent with the overdensity in Figure \ref{metals}. Furthermore, Figure
\ref{metals} also shows the derived [Fe/H], using the RAVE methodology, for
all stars from the M68 field. The overdensity centered on [Fe/H]$\sim-2$ are
the stars determined to be members of M68 in Paper II, and are well separated
in metallicity from the Galactic field stars. Using the equivalent width
method, in Paper II we calculated [Fe/H]$=-2.06\pm0.15$ for M68, again
consistent with that derived in this paper with the RAVE
method. \cite{Zwitter08} stated that it was not possible (or at least very
difficult) to provide a physical explanation for the calibration relation
between metallicities derived from equivalent widths or photometry methods to
those obtained by $\chi^2$ analysis. While this is still true, the current
paper provides further evidence that the calibration relations by
\cite{Zwitter08} do, in fact, hold.

\section{Conclusions}

We present for publication the stellar parameters of 29\,351 stars, observed
with AAOmega in fields centered on ten Galactic globular clusters. Parameters
were extracted using two pipelines, namely a version of the RAVE pipeline
adapted to work with AAOmega data and a pipeline based on the RAVE
methodology, and we publish here all parameters from both pipelines.

In addition, we find that the velocities, and velocity dispersions, of our
Galactic field stars (those not belonging to the clusters) in each field do
not agree well with those of the Besan\c{c}on Galaxy model. This descrepancy
may be due to the model lacking information on the many tidal features present
in the Galactic halo\new{, as well as nearby objects like the Magellanic
  Clouds}. It is apparent that there is a population which is kinematically
distinct from the Galaxy in the 47 Tuc field \new{(with $100\lesssim V_{\rm
    r}\lesssim200$\,km\,s$^{-1}$)}. Since \new{the SMC is in the background of
  this field, which has a radial velocity of $100\lesssim V_{\rm
    r}\lesssim200$\,km\,s$^{-1}$, it is clear that we are seeing SMC stars in
  this field.} We, therefore, suggest care is taken when using \new{the
  Besan\c{c}on} model for comparison with pencil beam surveys.

Furthermore, we have calculated calibrated metallicities ([M/H] and [Fe/H])
for each star based on the RAVE method outlined by \cite{Zwitter08} and these
correspond well with [Fe/H] calculated via the equivalent widths of the
calcium triplet lines originally published in Papers I, II and III.

\begin{acknowledgements}
This project has been supported by the University of Sydney, the Australian
Astronomical Observatory, the Australian Research Council, the Hungarian OTKA
grant K76816 and the Lend\"ulet Young Researchers Program of the Hungarian
Academy of Sciences. GyMSz acknowledges the Bolyai Fellowship of the Hungarian
Academy of Sciences. RRL acknowledges support from the Chilean Center for
Astrophysics, FONDAP Nr. 15010003, and from the BASAL Centro de Astrofisica y
Tecnologias Afines (CATA) PFB-06/2007. This publication was also financed by
the GEMINI-CONICYT Fund, allocated to project No. 32090010. We thank the
referee for their helpful comments and RRL would like to thank Martine
L. Wilson for her.
\end{acknowledgements}


\begin{thebibliography}{}

\bibitem[Barden et al.(2008)]{Barden08} Barden S.~C., Bland-Hawthorn
  J., Churilov V., et al., 2008, SPIE, 7014, 70144J

\bibitem[Baumgardt et al.(2010)]{Baumgardt10} Baumgardt H.,
  C{\^o}t{\'e} P., Hilker M., Rejkuba M., Mieske S., Djorgovski S.~G.,
  Stetson P., 2010, IAUS,  266, 365

\bibitem[Bekki \& Chiba(2002)]{Bekki02} Bekki K., Chiba M., 2002, ApJ,
  566, 245

\bibitem[Belokurov et al.(2007)]{Belokurov07} Belokurov V., Evans
  N.~W., Irwin M.~J., et al., 2007, ApJ, 658, 337

\bibitem[Cole et al.(2004)]{Cole04} Cole, A.~A., Smecker-Hane, T.~A.,
  Tolstoy, E., Bosler, T.~L., Gallagher, J.~S.\ 2004, MNRAS, 347,
  367

\bibitem[Conn et al.(2005)]{Conn05} Conn, B.~C., Lewis, G.~F., 
Irwin, M.~J., Ibata, R.~A., Ferguson, A.~M.~N., Tanvir, N., 
\& Irwin, J.~M.\ 2005, MNRAS, 362, 475 

\bibitem[Conn et al.(2007)]{Conn07} Conn, B.~C., et al.\ 2007, 
MNRAS, 376, 939 

\bibitem[Conn et al.(2008)]{Conn08} Conn B.~C., Lane R.~R., Lewis
  G.~F., Irwin M.~J., Ibata R.~A., Martin N.~F., Bellazzini M.,
  Tuntsov A.~V., 2008, MNRAS,  390, 1388

\bibitem[Evans \& Howarth(2008)]{Evans08} Evans, C.~J., \& Howarth,
  I.~D.\ 2008, MNRAS, 386, 826

\bibitem[De Propris et al.(2010)]{DePropis10} De Propris, R., Rich, R.~M.,
  Mallery, R.~C., \& Howard, C.~D.\ 2010, ApJL, 714, L249

\bibitem[Fr\'emat et al.(1996)]{Fremat96} Fr\'emat Y., Houziaux L.,
  Andrillat Y., 1996, MNRAS,  279, 25

\bibitem[Gebhardt \& Thomas(2009)]{Gebhardt09} Gebhardt, K., \&
  Thomas, J.\ 2009, ApJ, 700, 1690

\bibitem[Griffen et al.(2010)]{Griffen10} Griffen B.~F., Drinkwater
  M.~J., Thomas P.~A., Helly J.~C., Pimbblet K.~A., 2010, MNRAS, 405,
  375

\bibitem[Ibata et al.(1994)]{Ibata94} Ibata, R.~A., Gilmore, G., \& Irwin,
  M.~J.\ 1994, Nature, 370, 194

\bibitem[Kiss et al.(2007)]{Kiss07} Kiss, L.~L., Sz{\'e}kely, P., Bedding,
  T.~R., Bakos, G.~{\'A}., \& Lewis, G.~F.\ 2007, ApJL, 659, L129

\bibitem[Kissler-Patig et al.(1999)]{Kissler-Patig99} Kissler-Patig,
  M., Grillmair, C.~J., Meylan, G., Brodie, J.~P., Minniti, D., \&
  Goudfrooij, P.\ 1999, AJ, 117, 1206

\bibitem[Lane et al.(2009)]{Lane09} Lane R.~R., Kiss L.~L., Lewis
  G.~F., Ibata R.~A., Siebert A., Bedding T.~R., Sz{\'e}kely P.,
  2009, MNRAS, 400, 917 (Paper I)

\bibitem[Lane et al.(2010a)]{Lane10a} Lane R.~R., Kiss L.~L., Lewis
  G.~F., Ibata R.~A., Siebert A., Bedding T.~R., Sz{\'e}kely P.,
  2010a, MNRAS,  401, 2521 (Paper II)

\bibitem[Lane et al.(2010b)]{Lane10b} Lane R.~R., Kiss L.~L., Lewis
  G.~F., et al., 2010b, MNRAS, 406, 2732 (Paper III)

\bibitem[Lane et al.(2010c)]{Lane10c} Lane R.~R., Brewer B.~J., Kiss
  L.~L., et al., 2010c, ApJ,  711, L122

\bibitem[Lee et al.(2010)]{Lee10} Lee M.~G., Park H.~S., Hwang H.~S.,
  Arimoto N., Tamura N., Onodera M., 2010, ApJ,  709, 1083

\bibitem[Lipscy \& Plavchan(2004)]{Lipscy04} Lipscy S.~J., Plavchan
  P., 2004, ApJ,  603, 82

\bibitem[Mashchenko \& Sills(2005)]{Mashchenko05} Mashchenko S., Sills
  A., 2005, ApJ,  619, 258

\bibitem[Munari et al.(2005)]{Munari05} Munari, U., Sordo, R., Castelli, F.,
  \& Zwitter, T.\ 2005, A\&A, 442, 1127

\bibitem[Robin et al.(2003)]{Robin03} Robin A.~C., Reyl{\'e} C.,
  Derri{\`e}re S., Picaud S., 2003, A\&A, 409, 523

\bibitem[Reyl{\'e} et al.(2010)]{Reyle10} Reyl{\'e}, C., Robin, A.~C.,
  Schultheis, M., \& Marshall, D.~J.\ 2010, SF2A-2010: Proceedings of the
  Annual meeting of the French Society of Astronomy and Astrophysics.~Eds.:
  S.~Boissier, M.~Heydari-Malayeri, R.~Samadi and D.~Valls-Gabaud, p.51, 51

\bibitem[Scarpa et al.(2007)]{Scarpa07} Scarpa R., Marconi G.,
  Gilmozzi R., Carraro G., 2007, Msngr,  128, 41

\bibitem[Skrutskie et al.(2006)]{Skrutskie06} Skrutskie, M.~F., et al.\ 2006,
  AJ, 131, 1163

\bibitem[Sollima \& Nipoti(2010)]{Sollima10} Sollima A., Nipoti C.,
  2010, MNRAS,  401, 131

\bibitem[Storm et al.(2004)]{Storm04} Storm, J., Carney, B.~W., Gieren, W.~P.,
  Fouqu{\'e}, P., Freedman, W.~L., Madore, B.~F., \& Habgood, M.~J.\ 2004,
  A\&A, 415, 521

\bibitem[Warren \& Cole(2009)]{Warren09} Warren, S.~R., Cole, A.~A.\
  2009, MNRAS, 393, 272

\bibitem[Weldrake et al.(2004)]{Weldrake04} Weldrake, D.~T.~F., Sackett,
  P.~D., Bridges, T.~J., \& Freeman, K.~C.\ 2004, AJ, 128, 736

\bibitem[Zwitter et al.(2008)]{Zwitter08} Zwitter T., Siebert A.,
  Munari U., et al., 2008, AJ,  136, 421

\end{thebibliography}
\end{document}